# Tests of Sunspot Number Sequences: 3. Effects of Regression Procedures on the Calibration of Historic Sunspot Data

M. Lockwood[1] • M.J. Owens[1] • L. Barnard[1] •
I.G. Usoskin[2]

**Abstract**. We use sunspot-group observations from the Royal Greenwich Observatory (RGO) to investigate the effects of intercalibrating data from observers with different visual acuities. The tests are made by counting the number of groups [$R_B$] above a variable cut-off threshold of observed total whole-spot area (uncorrected for foreshortening) to simulate what a lower acuity observer would have seen. The synthesised annual means of $R_B$ are then re-scaled to the full observed RGO group number [$R_A$] using a variety of regression techniques. It is found that a very high correlation between $R_A$ and $R_B$ ($r_{AB} > 0.98$) does not prevent large errors in the intercalibration (for example sunspot-maximum values can be over 30 % too large even for such levels of $r_{AB}$). In generating the backbone sunspot number [$R_{BB}$], Svalgaard and Schatten (*Solar Phys.*, 2015) force regression fits to pass through the scatter-plot origin, which generates unreliable fits (the residuals do not form a normal distribution) and causes sunspot-cycle amplitudes to be exaggerated in the intercalibrated data. It is demonstrated that the use of Quantile-Quantile ("Q – Q") plots to test for a normal distribution is a useful indicator of erroneous and misleading regression fits. Ordinary least-squares linear fits, not forced to pass through the origin, are sometimes reliable (although the optimum method used is shown to be different when matching peak and average sunspot-group numbers). However, other fits are only reliable if non-linear regression is used. From these results it is entirely possible that the inflation of solar-cycle amplitudes in the backbone group sunspot number as one goes back in time, relative to related solar-terrestrial parameters, is entirely caused by the use of inappropriate and non-robust regression techniques to calibrate the sunspot data.

✉ M. Lockwood
   m.lockwood@reading.ac.uk

1  Department of Meteorology, University of Reading, UK

2  University of Oulu, Oulu, Finland

# 1. Introduction

Articles 1 and 2 of this series (Lockwood et al., 2016a; 2016b) provide evidence that the new "backbone" group sunspot number [$R_{BB}$] proposed by Svalgaard and Schatten (2015) overestimates sunspot numbers as late as Solar Cycle 17 and that this overestimation increases as one goes back in time. There is also some evidence that most of the overestimation grows in discrete steps, which could imply a systematic problem with the ordinary linear-regression techniques used by Svalgaard and Schatten to "daisy-chain" the calibration from modern values back to historic ones. This daisy-chaining is unavoidable in this context unless a method is used to calibrate historic (pre-photographic) data with modern data without relating both to data taken during the interim. (Note that one such a method, that avoids both regressions and daisy-chaining, has recently been developed by *Usoskin et al.* (2016).) As discussed in Articles 1 and 2, the regressions used are of particular concern because the daisy-chaining means that both random and systematic errors are amplified as one goes back in time.

As one reads the article by Svalgaard and Schatten (2015), one statement stands out and raises immediate concerns in this context: " Experience shows that the regression line almost always very nearly goes through the origin, so we force it to do so …" To understand the implications of this, consider two observers A and B, recording annual mean sunspot-group numbers $R_A$ and $R_B$, respectively. If observer B has lower visual acuity than A, then $R_B \leq R_A$. This may be caused by B having a lower resolution, and/or less well-focussed telescope, or one that gives higher scattered-light levels. It may also be caused by the keenness of observer B's eyesight and how conservative he/she was in making the subjective decisions to define spots and/or spot groups from what he/she saw. In addition, the local atmospheric conditions may also have hindered observer B (greater aerosol concentrations, more mists or thin cloud). Forcing the fits through the origin means that $R_A = 0$ when $R_B = 0$ and *vice-versa*. When the higher-acuity observer A sees no spot groups, the lower-acuity observer B should not detect any either and so both $R_A$ and $R_B$ should indeed both be zero in this case. However, there will, in general, have been times when observer A could detect groups but observer B could not and so $R_A > 0$ when $R_B = 0$. Thus any linear-regression fit used to scale $R_B$ to $R_A$ should not, in general, pass through zero as Svalgaard and Schatten (2015) forced all of their fits to do. There is no advantage gained by forcing the fits through the origin (if anything fits are easier to make without this restriction) but, as discussed in this article, it introduces the potential for serious error.

Figure 1a is a schematic that illustrates what effects this would have by plotting the variation of $R_B$ with $R_A$. The dot and dash line is the ideal case when observers A and B have the same visual acuity and are following the same rules to define spots and groups of spots so $R_B = R_A$. The solid line in (a) is when observer B has lower acuity and so $R_A > R_B$ but the variation of $R_B$ with $R_A$ remains linear. As discussed above, in general $R_A > 0$ when $R_B = 0$. The dashed line is the best fit linear regression that is forced to pass through the origin. The horizontal lines demonstrate how scaling off a value for $R_A$ from measured values of $R_B$ using this regression fit will cause an underestimate of $R_A$ at lower-than-average values but an overestimate at higher-than-average values. Hence the amplitude of solar cycles is falsely amplified by the assumption that $R_A = 0$ when $R_B = 0$ and forcing the fits through the origin of the regression plot. Given the high correlations and the similar appearance of the various regression lines, it would be easy to dismiss this effect as small; however, we here use the distributions of whole spot group areas from the Royal Greenwich Observatory (RGO) group area data (Willis et al., 2013) to show that it can be a highly significant effect, especially when one considers that the effect will be compounded by successive intercalibrations in the daisy chain.

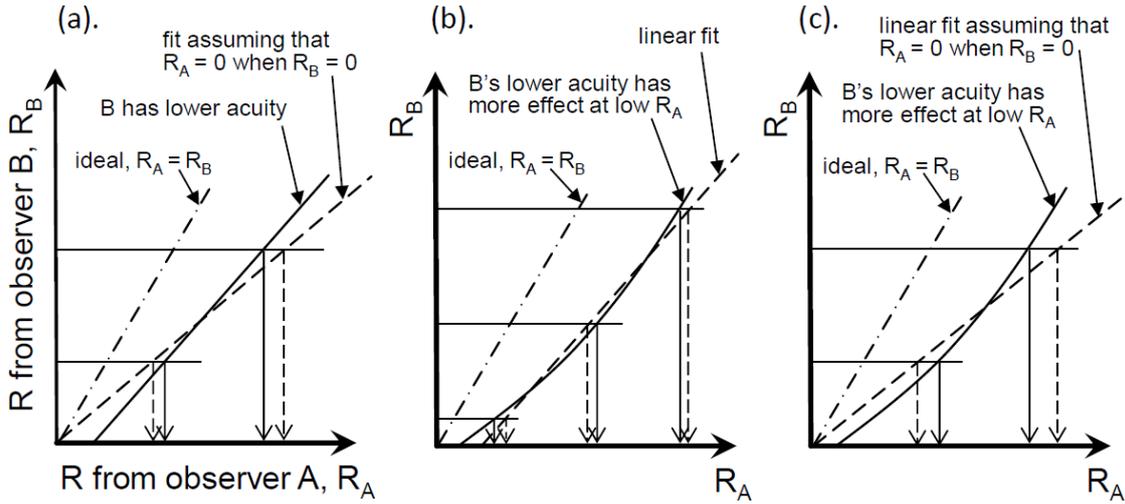

**Figure 1**. Schematics illustrating the importance of the intercept and linearity when using a linear regression fit to intercalibrate sunspot group numbers seen by an observer B [$R_B$] to evaluate what observer A would have seen [$R_A$] when observer B has lower visual acuity. See text for details.

Other concerns are that the errors in the data do not meet the requirements set by the assumptions of ordinary least-squares (OLS) fitting algorithms, and this possibility should always be tested for using the fit residuals. Failure of these tests means an inappropriate fitting procedure has been used or the noise in the data is distorting the fit. In addition, OLS can be applied by minimising the perpendiculars to the best-fit line or by minimising the verticals to the fit line. It can be argued that this choice should depend on the relative magnitudes of the errors in the fitted parameters. Another possibility that we consider here is that the effect of reduced acuity of observer B may vary with the level of solar activity leading to non-linearity in the relationship between $R_A$ and $R_B$ (see Usoskin et al., 2016 for evidence of this effect). We here also investigate the effects of using the linear ordinary least squares fits used by Svalgaard and Schatten (2015) under such circumstances.

Figure 1b illustrates the effects of using a linear fit if observer B's lower acuity has more effect at low sunspot numbers than at high ones, giving a non-linear (quadratic) relationship. In this case, a linear regression with non-zero intercept causes inflation of both the highest and the lowest values but lowers those around the average. Figure 1c shows the effects of both using a linear fit and making it pass through the origin, as employed by Svalgaard and Schatten (2015): in this case the effects are as in Figure 1a but the non-linearity makes them more pronounced.

Non-linearity between the two variables is just one of the main pitfalls in OLS regression. These can arise because the data violate one of the four basic assumptions that are inherent in the technique and that justify the use of linear regression for purposes of inference or prediction. The other pitfalls are a lack of statistical independence of the errors in the data; heteroscedasticity in the errors (they vary systematically with the fit parameters); and cases for which the errors are not normally distributed (about zero). In particular, one or more large-error datapoints can exert undue "leverage" on the regression fit. If one or more of these assumptions is violated (*i.e.* if there is a nonlinear relationship between the variables or if their errors exhibit correlation, heteroscedasticity, or a non-Gaussian distribution) then the forecasts, confidence intervals, and scientific insights yielded by a regression model may be seriously biased or misleading. If the fit is correct, then the fit residuals will reflect the errors in the data and so we can apply tests to the residuals to check that none of the assumptions has been invalidated. Non-linearity is often evident as a systematic pattern when one plots the fit residuals against either of the regressed variables. For regression of time-series data, lack of independence of the errors is seen as high persistence of the fit residual time series. Lack of homoscedasticity is apparent from scatter plots because the scatter increases systematically with the variables. A normal distribution of fit residuals can be readily tested for using a Quantile–Quantile ("Q – Q") plot (*e.g.* Wilk and Gnanadesikan, 1968). This is a graphical technique for determining if two data sets come from populations with a common distribution; hence by making one of the datasets normally-distributed we can test the other to see if it also has a normal distribution. Erroneous outliers and lack of linearity can also be identified from such Q – Q plots. If outliers are at large or small values they can have a very large influence on a linear regression fit – such points can be identified because they have a large Cook's-D (leverage) factor (Cook, 1977) and should be removed and the data re-fitted. There is no one standard approach to regression that can be applied and implicitly trusted. There are many options that must be investigated, and the above tests must be applied to ensure that the best option is used and that the results are statistically robust. In addition to OLS, we here employ non-linear regression (using second-order and third-order polynomials), median least squares (MLS) and bayesian least squares (BLS). The MLS and BLS procedures were discussed by Lockwood et al. (2006).

The results presented in this article show that linear regression fits in the context of intercalibrating sunspot-group numbers can violate the inherent assumptions and lead to some very large errors, even though the correlation coefficients are high. In Section 2, we present one example in which inter-calibration over 2 full sunspot cycles (1953 – 1975) can produce an inflation of sunspot peak values of over 30 % even when the correlation between $R_A$ and $R_B$ exceeds 0.98. This is a significant error. To put it into some context, Svalgaard (2011) pointed out a probable discontinuity in sunspot numbers around 1945 that has been termed the "Waldmeier discontinuity". Svalgaard quantified it as a 20 % change but Lockwood, Owens and Barnard (2014) and Lockwood et al. (2016a) find it is 11.9±0.6 % and Lockwood, Owens, and Barnard (2016) find it to be 10 %. (The latter estimate is lower because it is the only one not to assume proportionality). Hence 30 % is a very significant number for one intercalibration, let alone when it is combined with the effect of others in a series of intercalibrations. In Section 3 we present a second example interval (1923 – 1945, when solar activity was lower) to see if it reveals the same effects.

Lastly, we note that we here employ annual means to be consistent with Svalgaard and Schatten (2015). We do not test for any effects of this in the present article but it does cause additional concerns when the data are sparse. This is because observers A and B may have been taking measurements on different days and, because of factors such as regular annual variations in cloud obscuration, their data could even mainly come from different phases of the year. This may therefore not be a random error, which would again invalidate the assumptions of ordinary least squares regression. Usoskin et al. (2016) show this effect can be highly significant for sparse data and Willis, Wild, and Warburton (2016) show it even needs to be considered when using the earliest (before 1885) data from the Royal Observatory, Greenwich.

In the present article, we make use of the photoheliographic measurements from the Royal Observatory, Greenwich and the Greenwich Royal Observatory (here collectively referred to as the "RGO" data). We employ the version of the RGO data made available by the Space Physics website (solarscience.msfc.nasa.gov/greenwhch.shtml) of the Marshall Space Flight Center (MSFC) which has been compiled, maintained and corrected by D. Hathaway. These data were downloaded in June 2015. As noted by Willis et al. (2013b), there are some small differences between these MSFC data and versions of the RGO data stored elsewhere (notably those in the National Geophysical Data Center, NGDC, Boulder: www.ngdc.noaa.gov/nndc/struts/results?op_0=eq&v_0=Greenwich&t=102827&s=40&d=8&d=470&d=9). We here use only data for 1923 – 1976 for which these differences are minimal. The use of this interval also avoids all times when the calibration of the RGO data has been questioned (Cliver and Ling, 2016; Willis, Wild, and Warburton, 2016).

## 2. Study of 1953 -1975

### 2.1 Distribution of Sunspot Group Areas

Figure 2 shows the distribution of whole spot areas [$A$] in each defined sunspot group from the RGO data (Willis et al., 2013a) in the interval 1953-1975. $A$ is uncorrected for fore-shortening and so is the area that the observer actually sees on the solar disc. The-right hand plot is a detail of the left-hand plot and shows the peak of the distribution. The large number of small-area groups mainly arises from near the solar limb where the foreshortening effect is large. These areas are those recorded by the RGO observers, who are here collectively termed "Observer A". To simulate what a lower-acuity Observer "B" would have seen, we here assume that he/she would only detect groups for which the observed area [$A$] exceeded a threshold [$A_{th}$]. The number of groups seen on each day by the RGO observers and by the virtual observer B [$R_A$ and $R_B$ respectively] were counted. Annual means of both $R_A$ and $R_B$ were then evaluated to be compatible with the procedure used to generate the backbone data series [$R_{BB}$]. This was repeated for a wide range of $A_{th}$ thresholds.

### 2.2. Variations of R$_A$ and R$_B$ and Fits of R$_B$ to R$_A$

Figure 3 shows the variations of observed $R_A$ (black line) and synthesised $R_B$ (dashed line) sunspot-group numbers for this interval, for an example value of $A_{th}$ of $150 \times 10^{-6} A_{SH}$ (where $A_{SH}$ is the area of a solar hemisphere). The coloured lines are the best fits of $R_B$ to $R_A$ using various regression procedures detailed in Table 1. The fits for MLS and BLS (fit 6 and fit 7, respectively) are not shown because the results are almost identical to those for fit 2 because the scatter in the data is low. The fit for the third-order polynomial (fit 8) is not shown because the Q − Q plot reveals it to be less robust than that for the second-order polynomial (see below). Note that the larger of the two peaks in Figure 3 is overestimated by most of the fits, but particularly by fits 3 and 4, which force the regression line to pass through the origin. These fits also underestimate the sunspot-minimum values, as was predicted by Figure 1. However, the linear fits that are free to determine the intercept also overestimate the largest values (albeit to a smaller degree).

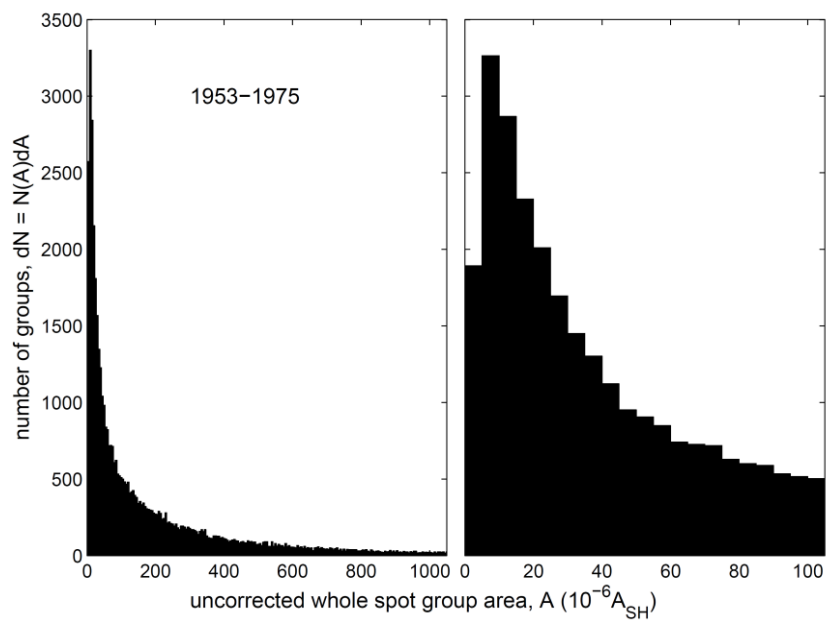

**Figure 2**. Distribution of uncorrected whole-spot sunspot-group areas [$A$] measured by RGO observers over the interval 1953-1975 for bin widths d$A$ = $(5 \times 10^{-6}) A_{SH}$, where $A_{SH}$ is the area of a solar hemisphere. The distribution on the right is a detail of that on the left showing the peak near $A$=0 more clearly. Note that $A$ is uncorrected for fore-shortening effect near the solar limb and so is the area actually seen by the observer.

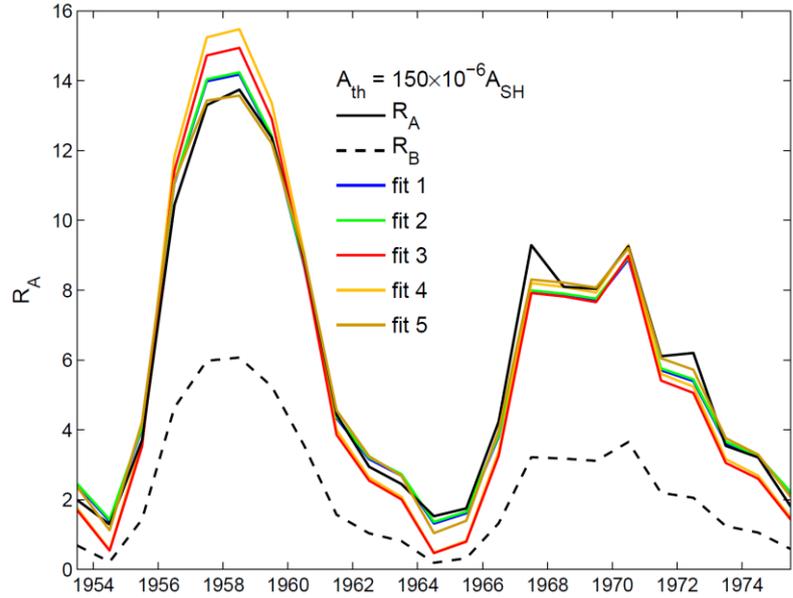

**Figure 3.** Time series of observed and fitted sunspot-group numbers for the interval 1953-1975. The black line is the number of groups [$N_A$] detected by the RGO observers, who are here termed observer A. The dashed line is the number [$N_B$] that would have been detected by lower-acuity observer B if he/she could only detect groups with uncorrected (for foreshortening) whole spot area $A > A_{th} = 150 \times 10^{-6} A_{SH}$ where $A_{SH}$ is the area of a solar hemisphere. The coloured lines show the results of different fits of $N_B$ to $N_A$ described in the text (see Table 1).

**Table 1.** Fit procedures employed

| Fit | Line colour in Figures | Fit type | Assumed variation | Parameter minimised | Treatment of intercept |
|---|---|---|---|---|---|
| 1 | Blue | OLS | linear | r.m.s. of perpendiculars | Not forced through origin |
| 2 | Green | OLS | linear | r.m.s. of verticals | Not forced through origin |
| 3 | Red | OLS | linear | r.m.s. of perpendiculars | Forced through origin |
| 4 | Orange | OLS | linear | r.m.s. of verticals | Forced through origin |
| 5 | Brown | Poly-nomial | 2$^{nd}$-order polynomial | r.m.s. perpendiculars | Not forced through origin |
| 6 | - | MLS | linear | r.m.s. perpendiculars | Not forced through origin |
| 7 | - | BLS | linear | r.m.s. perpendiculars | Not forced through origin |
| 8 | Cyan | Poly-nomial | 3$^{rd}$-order polynomial | r.m.s. perpendiculars | Not forced through origin |

The scatter plot of annual means of $R_B$ as a function of $R_A$ (for this $A_{th}$ of 150 × 10$^{-6}A_{SH}$) is shown in Figure 4 along with the best-fit regression lines used to derive the variations shown in Figure 3. Figure 5 presents the Q – Q plots of the ordered normalised fit residuals against predictions for a normal distribution for the OLS fits 1 - 5 and 8. For valid OLS fits, the residuals should be normally distributed, which would make all points lie on the diagonal line in the Q – Q plot. (If the points were to lie on a straight line but the slope was not unity, it would mean that one or both regressed parameters have a distribution with a different kurtosis (sharpness of peak) compared to a Gaussian; if the points were to form a characteristic S-shape about the origin it would mean that there is a skewness in one or both distributions; if the variation were to be complex it would mean that there are a major deviations from the assumptions of the regression). Figure 5 shows that none of fits 1-4 pass this test, and so these fits are unreliable and should not be used. The fits that have forced the regression line through the origin (fits 3 and 4) are particularly poor. However, fits 1 and 2 are also not ideal, in neither the Q − Q plots shown in Figure 5 (particularly the tails of the distributions) nor in the magnitude of the largest fitted values in Figure 3. The Q − Q plot for the second-order polynomial (fit 5, Figure 5e) is closest to ideal (especially when the worst outlier is omitted, see below), but the tails of the distribution become non-Gaussian again when a third-order polynomial was used (fit 8, Figure 5f).

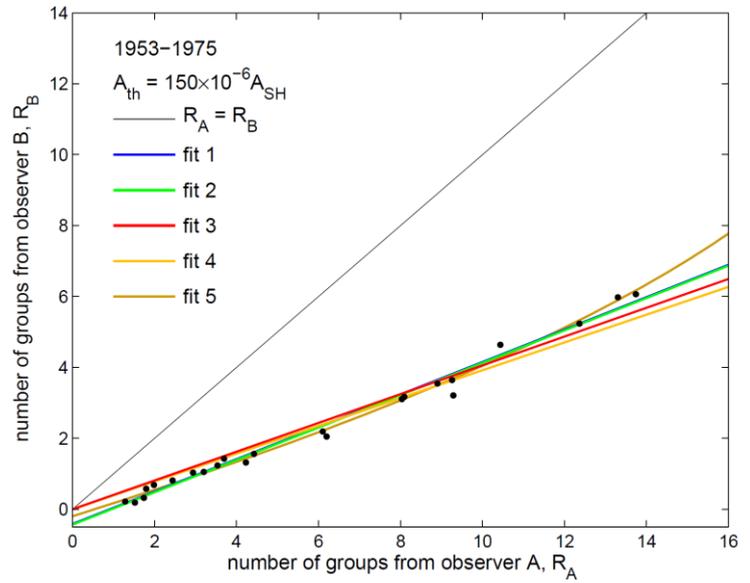

**Figure 4.** Scatter plot of $R_B$ as a function of $R_A$ (black points) for 1953-1975 and the same area threshold $A_{th} = 150\times10^{-6}A_{SH}$ as used in Figure 3. The coloured lines are the best-fit regression lines (shown using the same colour scheme as Figure 3 and given in Table 1.

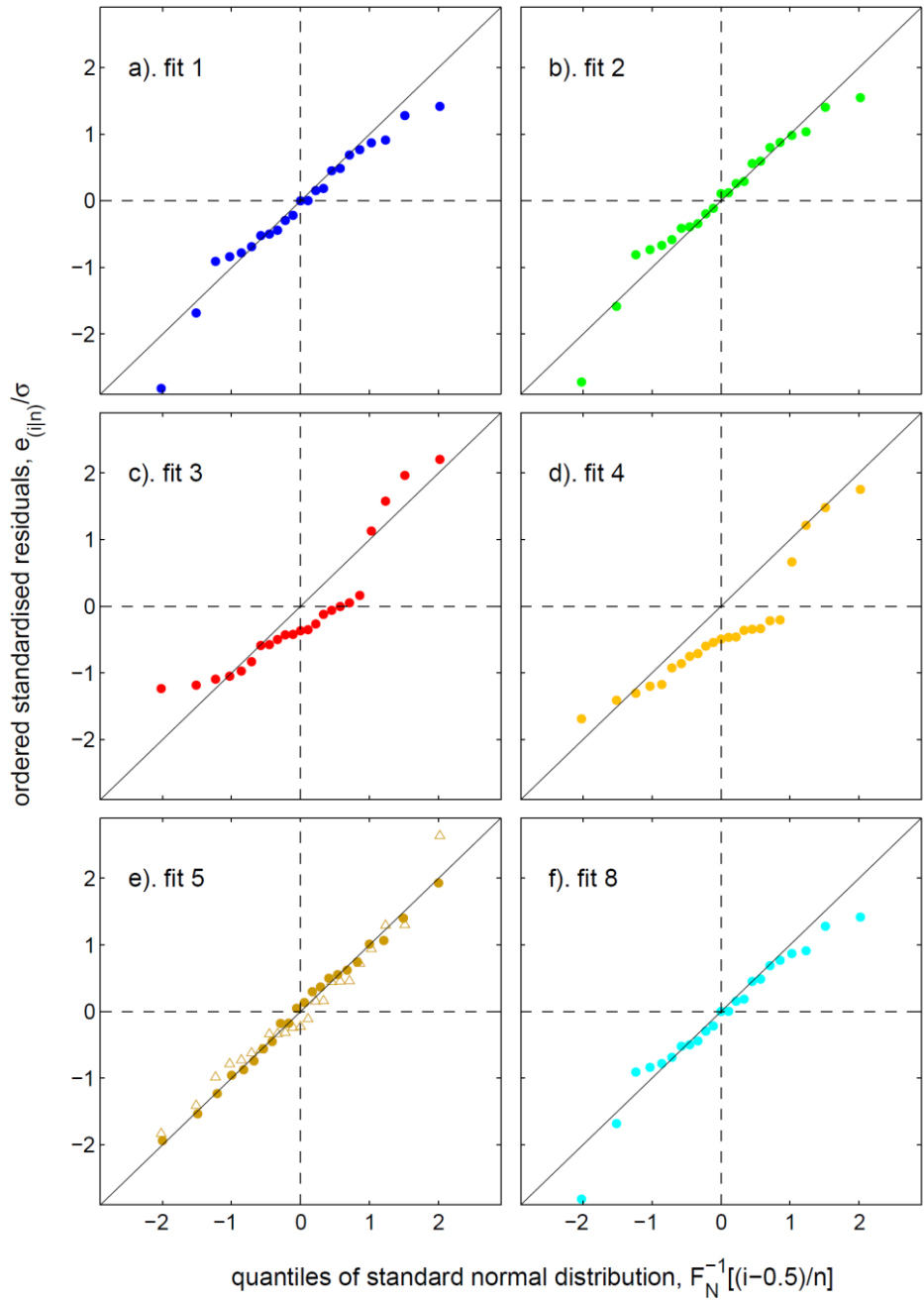

**Figure 5**. Q – Q plots of ordered normalised fit residuals against predictions for a normal distribution for fits 1-5 and 8 for 1953-1975. For valid OLS fits the residuals must be normally distributed which would make all points lie on the diagonal line. In (e) for fit 5, the open triangles show the results for all datapoints, whereas the solid circles are after removal of the largest outlier.

Fits using median least squares (MLS, fit 6) and Bayesian least squares (BLS, fit 6) were also made but were no better than the comparable OLS fit (fit 1). We also attempted successive removal of the largest outliers to try to make the fits converge to a stable result, but again no improvement was made for all these linear fits. This left just one assumption to test, namely that the variation of $R_B$ with $R_A$ is linear. A least-squares fit of a second-order polynomial fit was carried out (fit 5): this is shown by the brown lines in Figures 3 and 4. This appears to remove the problem of the exaggerated peak values. Note that for this fit one outlier data point has been removed (see below). In addition, a $3^{nd}$-order polynomial fit was carried out (fit 8): the Q − Q plot for this fit is shown by the cyan points in Figure 5f, and it can be seen that this fit generates some non-Gaussian tails to the distribution.

In Figure 5e, the open triangles show the results for the second-order polynomial fit to all datapoints and the point in the upper tail of the distribution is seen to be non-Gaussian. This arises from the outlier data point that can be seen in Figure 3 at $R_A \approx 9.3$, $R_B \approx 3.2$. The solid circles are for the fit after this outlier has been removed and the remaining points can now be seen to give an almost perfect Gaussian distribution of residuals, and so the fit is robust. The brown lines in Figures 3 and 4 show the results of this fit with the outlier removed. The largest outlier was also removed or all other fits but fit 5 was the only one for which the Q − Q plot was significantly improved. Note that for the test done here, the fits are never used outside the range of values that were used to make the fit. However, this would not necessarily be true of an intercalibration between two daisy-chained data segments and very large errors could occur if there is non-linearity and one is extrapolating to values outside the range used for calibration fitting.

## 2.3. Effect of the Threshold $A_{th}$

The results in the last section were all for a single example value of the observable area threshold for observer B: $A_{th} = 150 \times 10^{-6} A_{SH}$. To study the effect of various levels of visual acuity of observer B, we here vary $A_{th}$ between 0 (for which $R_B = R_A$) and $650 \times 10^{-6} A_{SH}$ in steps of $5 \times 10^{-6} A_{SH}$ for the same dataset (1953 – 1975). For each $A_{th}$ we compute: the correlation coefficient $r_{AB}$ between $R_B$ and $R_A$; the percentage by which the fitted peak value in the interval is greater than the corresponding peak of $R_A$ [$\Delta_{peak}$] and the percentage by which the average of the fitted value in the interval is greater than the average of $R_A$ [$\Delta_{mean}$]. The results are shown here for fits 1, 2, 3, 4, and 5 in Figure 6. The top panel shows that the correlation remains extremely high ($r_{AB} > 0.97$) even for very large $A_{th}$ of $650 \times 10^{-6} A_{SH}$. However, these high correlations do not prevent considerable errors in the fitted mean and peak values arising from the regression procedure. The fits that force the regression through the origin give very large errors (fit 3 overestimating the peak value by 15% and fit 4 by 30% at $A_{th} = (450 \times 10^{-6}) A_{SH}$, for which $r_{AB}$ still exceeds 0.98. Hence high correlation is certainly no guarantee of reliable regression fits. The mean values tend to be decreased because of the fall in the low values expected from Figure 1; however, for very large $A_{th}$, fit 3 yields increases in both mean and peak values. Fit 2 does not change the mean values at all but it does inflate the peak values as much as does fit 1. The only fit not to do this is the non-linear fit 5 which does slightly decrease the peak values, particularly at the larger $A_{th}$.

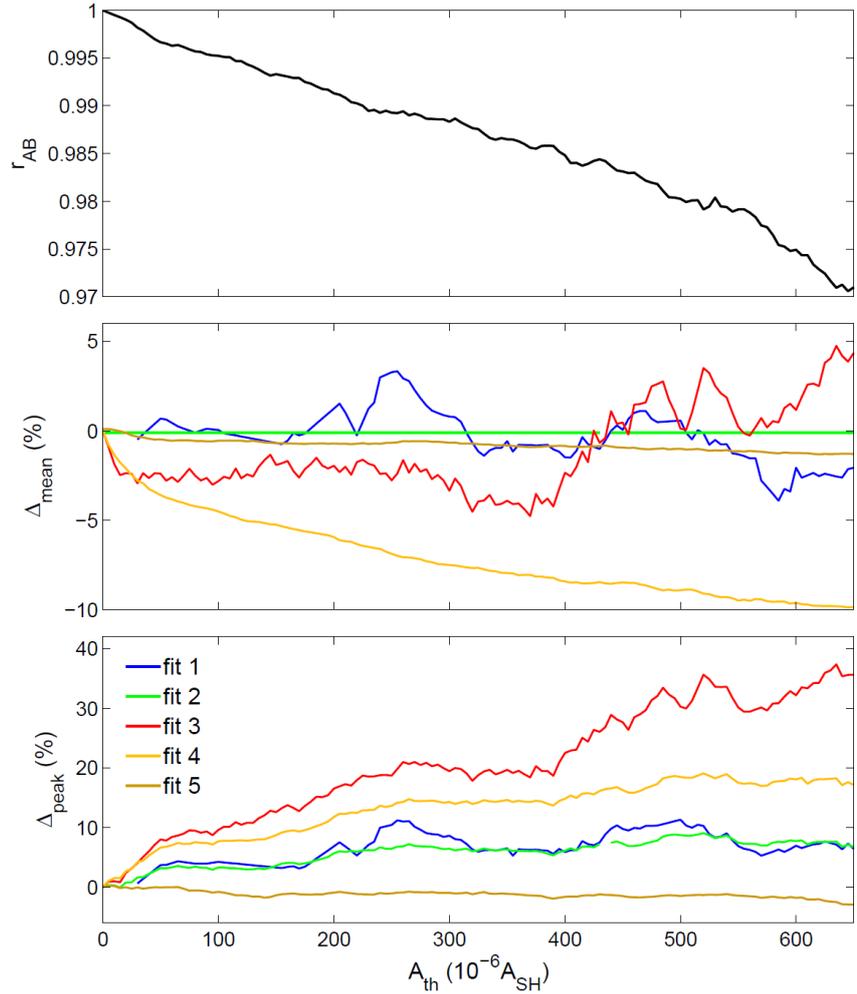

**Figure 6.** The effect of varying the threshold $A_{th}$ for the 1953-1975 data subset. (top) $r_{AB}$, the correlation between $R_B$ and $R_A$; (middle) $\Delta_{mean}$, the percentage by which the average of the fitted value in the interval is greater than the average of $R_A$ and (bottom) $\Delta_{peak}$, the percentage by which the fitted peak value in the interval is greater than the peak of $R_A$. Coloured lines are as in Figures 3 and 4 and as given by Table 1.

Hence of all the fits, for this calibration interval, the non-linear fit 5 performs best, but it is still not ideal. Looking at Figure 4 we see that the non-linearity is very subtle indeed and not at all obvious from the scatter plot, but our tests show it is an important factor in this case.

## 3. Study of 1923 – 1945

It is found that the effects, in general, depend on both the length of the regression interval and which interval is chosen. There are too many combinations of possibilities to attempt a parametric study, but we here chose a second interval of length two solar cycles (1923-1945) which illustrates somewhat different behaviour from the last section. Figure 7 corresponds to Figure 2. It shows the form of the distribution of observed areas [A] is the same for the two intervals, but the numbers in each bin are lower for 1923 – 1945 because solar activity is lower. Figures 8, 9, and 10 show, respectively, the time series and fits, the scatter plots with regression lines, and the Q – Q plots for this interval, in the same formats as Figures 3, 4, and 5, respectively. The difference between the fits is almost undetectable in Figure 9 but Figure 10 shows that the Q – Q plots still violate the required normal distribution of residuals for the fits that force the regression through the origin (fits 3 and 4). For this interval the regressions that are not forced through the origin (fits 1 and 2) give good alignment along the diagonal line and hence are robust fits with a normal distribution in the fit residuals. In this case the polynomial fits also produce very good normal distributions on the Q – Q plot because the best-fit higher-order coefficients of the polynomial are very close to zero, and the best fit is essentially linear.

**Figure 7.** Distribution of uncorrected whole spot sunspot group areas [$A$] measured by RGO observers over the interval 1923-1945 for bin widths $dA=(5\times10^{-6})A_{SH}$ where $A_{SH}$ is the area of a solar hemisphere. The distribution on the right is a detail of that on the left showing the peak near $A=0$ more clearly. Note that $A$ is uncorrected for foreshortening effect near the solar limb and so is the area actually seen by the observer.

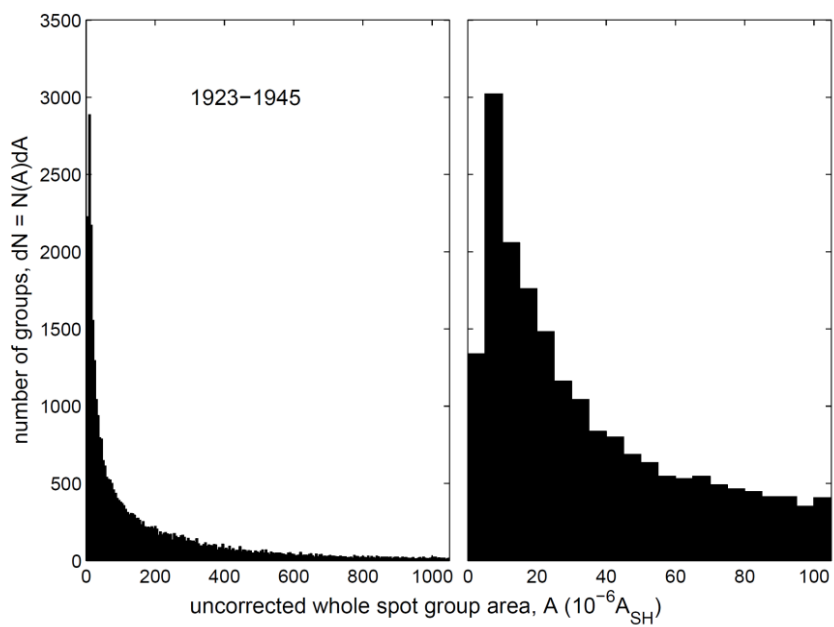

**Figure 8.** Time series of observed and fitted group sunspot numbers for the interval 1923-1945. The black line is the number of groups [$N_A$] detected by the RGO observers, who are here termed observer A. The dashed line is the number [$N_B$] that would have been detected by lower-acuity observer B if he/she could only detect groups with uncorrected (for foreshortening) whole spot area $A > A_{th} = 150\times10^{-6}A_{SH}$, where $A_{SH}$ is the area of a solar hemisphere. The coloured lines show the results of different fits of $N_B$ to $N_A$ described in the text (see Table 1).

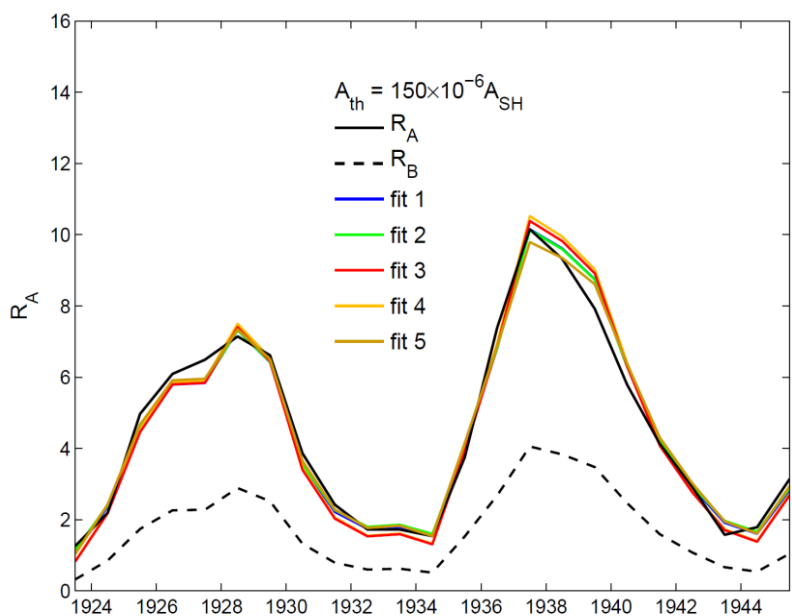

**Figure 9.** Scatter plot of $R_B$ as a function of $R_A$ (black points) for 1923-1945 and the same area threshold $A_{th} = 150\times10^{-6}A_{SH}$ as used in Figure 8. The coloured lines are the best-fit regression lines (shown using the same colour scheme as Figure 8 and given in Table 1).

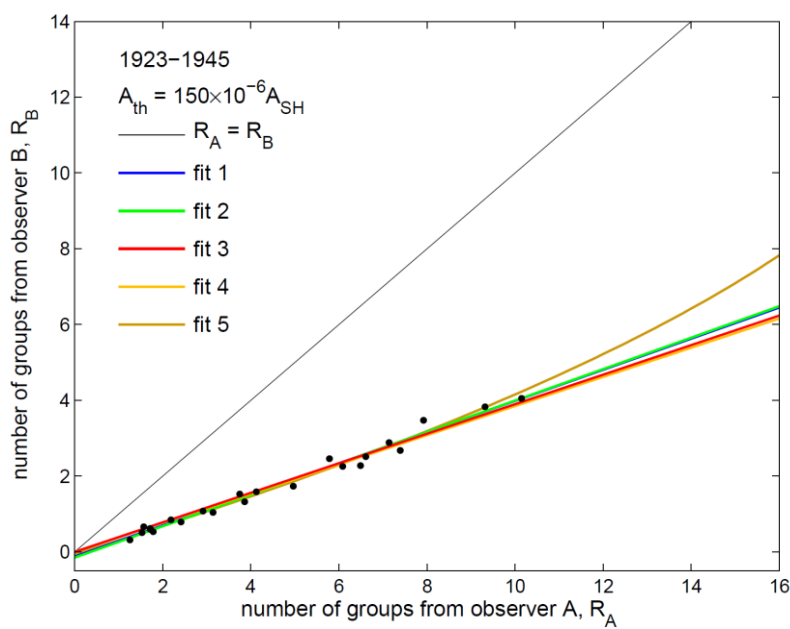

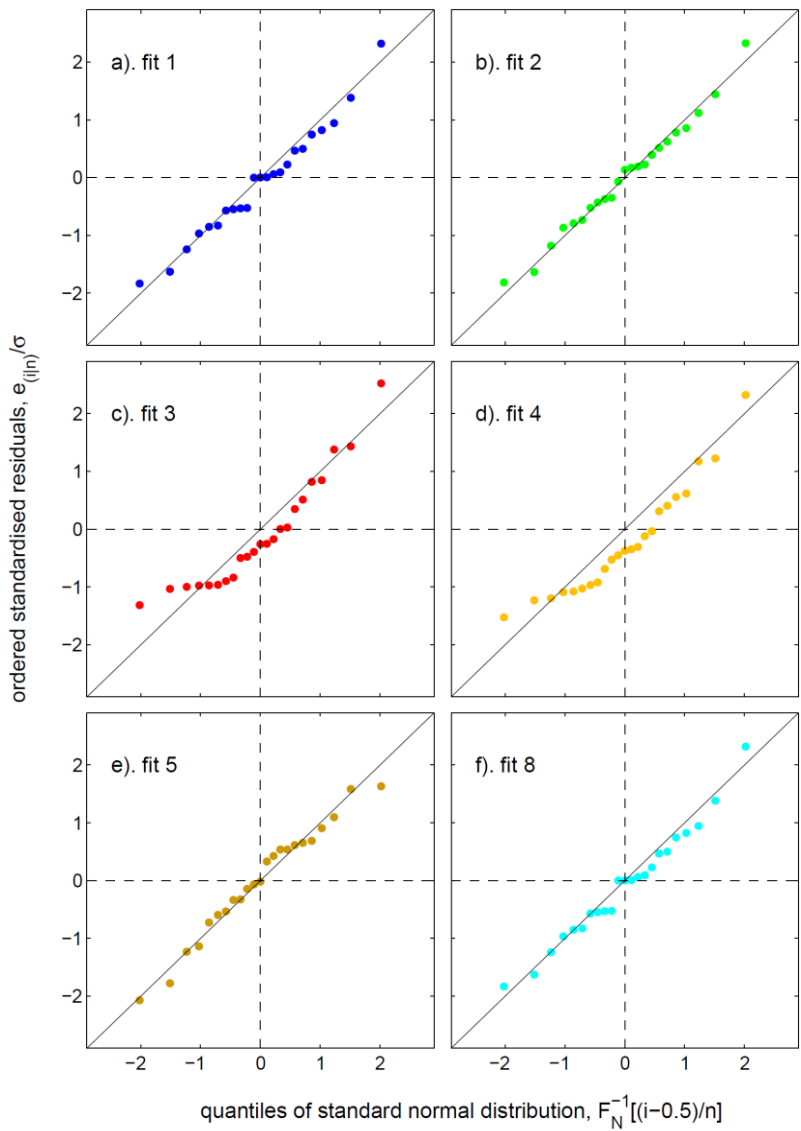

**Figure 10**. Q – Q plots of ordered normalised fit residuals against predictions for a normal distribution for fits 1-5 and 8 for 1923-1945. For valid OLS fits, the residuals must be normally distributed, which would make all points lie on the diagonal line. In this case there are no clear outliers that require removal.

Figure 11 corresponds to Figure 6: there are some similarities but there are also significant differences. For this 1923 – 1945 example, fit 2 is the closest that we have obtained to being ideal, there is no change in the mean value at any $A_{th}$ (as for the previous example) and the percentage changes in the peak value are much smaller than we have obtained before. Remember that fit 2 assumes that observer A's data are correct and all of the uncertainty is in observer B's data (because it minimise the r.m.s. of the verticals in the scatter plot). This is perhaps what we would have expected given the uncertainties in the lower-acuity data should be larger than those in the higher acuity data. However fit 1 also performs quite well with relatively small errors in the mean values and almost none in the peaks. Again this appears to be consistent with expectations as this fit minimises the perpendiculars, which should be a better thing to do when the uncertainties in $R_A$ and $R_B$ are comparable and this is more likely to be the case when solar activity is high. The non-linear fit again gives almost no error in mean values but persistently tends to underestimate the peak values in this case.

## 4. Discussion and Conclusions

Our tests of regression procedures, comparing the original RGO sunspot group area data with a deliberately degraded version of the same data, show that there is no one definitive method that ensures the regressions derived are robust and accurate. Certainly correlation coefficient is not a valuable indicator and very high correlations are necessary but very far from sufficient.

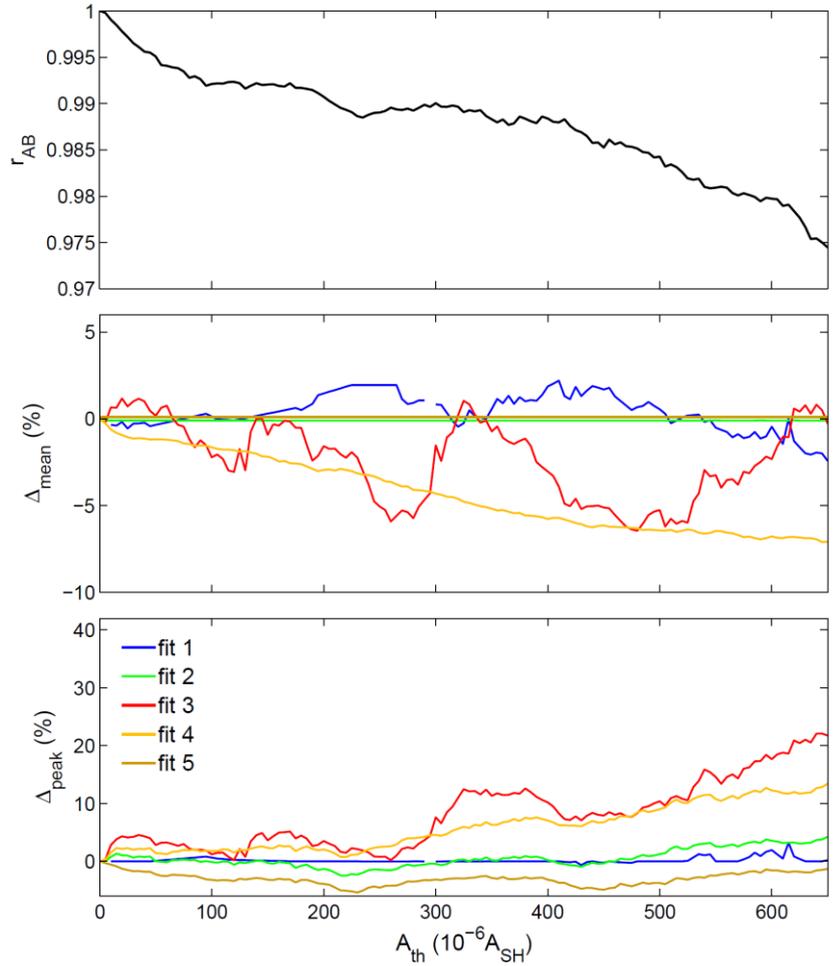

**Figure 11.** The effect of varying the threshold $A_{th}$ for the 1923-1945 data subset. (top) $r_{AB}$, the correlation between $R_B$ and $R_A$; (middle) $\Delta_{mean}$, the percentage by which the average of the fitted value in the interval is greater than the average of $R_A$ and (bottom) $\Delta_{peak}$, the percentage by which the fitted peak value in the interval is greater than the peak of $R_A$. Coloured lines are as in previous figures and as given by Table 1.

The one definitive statement that we can make is that forcing fits through the origin is a major mistake. It causes solar-cycle amplitudes to be inflated so that peak values in the lower acuity data are too high and both minimum and mean values are too low. This is the method used by Svalgaard and Schatten (2015), and our findings show that it will have contributed to a false upward drift in their backbone group number reconstruction [$R_{BB}$] values as one goes back in time. At the time of writing we do not have the original data to check the effect on both the regressions used to intercalibrate backbones and any regressions used to combine data into backbones. Both will be subject to this effect. Hence we cannot tell whether or not this explains all of the differences between, for example, the long term changes in $R_{BB}$ and the terrestrial data (ionospheric, geomagnetic, and auroral) discussed in Articles 1 and 2. However, it will have contributed to these differences. Note that all of the above also applies to any technique based on the ratio $R_A/R_B$ as that also forces the fit through the origin.

Lastly it is not clear which procedure should be used to daisy-chain the calibrations. Ordinary least-squares fits work well only when the Q–Q plots show a good normal distribution of residuals. Even then, minimising the verticals gives the best answer for the mean values but minimising the perpendiculars gives the best answer for the peak values. The failures in the Q − Q plots appear to be mainly because the dependence is not linear and a non-linear fit then works well. We used a second-order polynomial and the fitted $R_B^2$ term is found to be relatively small (meaning it is a near-linear fit) and hence this seems to have been adequate, at least for the cases we studied. However, we note that this should not be used for values that are outside the range seen during the inter-calibration interval because the dependence of the extrapolation on the polynomial used is then extremely large.

**Acknowledgements** The authors are grateful to David Hathaway and the staff of the Solar Physics Group at NASA's Marshall Space Flight Center for maintaining the on-line database of RGO data used here The work of M. Lockwood, M.J.Owens and L.A.Barnard at Reading was funded by STFC consolidated grant number ST/M000885/1 and that of I.G.Usoskin was done under the framework of the ReSoLVE Center of Excellence (Academy of Finland, project 272157).